\documentclass{aastex}
\def\ls{{_<\atop^{\sim}}}
\def\gs{{_>\atop^{\sim}}}

\def\lax    {${_<\atop^{\sim}}$}
\def\gax    {${_>\atop^{\sim}}$}

\slugcomment{Submitted to ApJ}
\shorttitle{}
\shortauthors{\m}

\begin{document}
\small
\title{The complex X-ray absorbers of NGC~3516 observed by BeppoSAX}
\author{E.~Costantini$^{1}$, F.~Nicastro$^{1,2,3}$, A.~Fruscione$^{1}$, 
S.~Mathur$^4$, A.~Comastri$^{5}$, M.~Elvis$^{1}$, F.~Fiore$^{2}$, 
 C.~Salvini$^5$, G.M.~Stirpe$^{5}$, C.~Vignali$^{5,6}$, B.~Wilkes$^{1}$, P.T.~O'Brien$^{7}$ \& M.R.~Goad$^{7}$}

\affil{$^{1}$ Harvard-Smithsonian Center for Astrophysics, 60 Garden Street, 
Cambridge, MA 02138, USA}
\affil{$^2$ Osservatorio Astronomico di Monteporzio, Via Osservatorio 2, 
I-00040, Monteporzio, Italy} 
\affil{$^3$ Istituto di Astrofisica Spaziale - CNR, Via del Fosso del 
Cavaliere, Roma, I-00133, Italy}
\affil{$^4$ Ohio State University, Dept. of Astronomy, 174 W. 18th Ave., 
Columbus, OH 43210-1106}
\affil{$^5$ Osservatorio Astronomico di Bologna, Via Ranzani 1, I-40127, 
Bologna, Italy}
\affil{$^6$ Dipartimento di Astronomia, Universit\`a di Bologna, Via
Ranzani 1, 40127, Bologna, Italy}
\affil{$^7$ Department of Physics and Astronomy, University of Leicester, 
University Road, Leicester, LE1 7RH, England, UK}

\begin{abstract}
In this paper we present the analysis of two broad band 
(0.1-150 keV) BeppoSAX observations of the Seyfert~1 galaxy 
NGC~3516. The two observations were taken 4 months apart, on 
1996 November 8 and 1997 March 12 respectively. 
We report a dramatic change in the degree of obscuration of the 
central source between the two observations, and propose, as possible explanations, transient 
absorption by either a stationary-state cloud of cold gas 
crossing the line of sight, or a varying-state, initially neutral and 
dense amount of expanding gas with decreasing density and therefore decreasing opacity. 

We also report the detection of a second highly ionized absorber/emitter, which causes deep FeXVII-XXII K edges at $\sim 7.8$ keV to appear in both 
BeppoSAX spectra of NGC~3516, and possibly produces the soft X-ray continuum 
emission and the 1 keV blend of Fe L recombination lines detected 
during the epoch of heavy nuclear obscuration. 

\end{abstract}

\keywords{galaxies: Seyfert -- galaxies: individual
(NGC~3516) -- X-ray: galaxies }
\section{ Introduction}

NGC~3516 is a bright (F$_{2-10 keV} = 7.8 \times 10^{-11}$ erg 
cm$^{-2}$ s$^{-1}$, Reynolds 1997) nearby (z = 0.009, D = 54 Mpc, 
for $H_0 = 50$ km $s^{-1}$ Mpc$^{-1}$) Seyfert 1 galaxy which is 
known to host complex and variable systems of both mildly and highly 
ionized absorbers in the UV and in the X-rays. 
In the UV band, IUE spectra of NGC~3516 taken between 1978 June and 1989 
October (Ulrich and Boisson 1983, Voit et al. 1987, Walter et al. 1990, 
Kolman et al. 1993) show the presence of at least two systems of CIV
resonant absorption lines: one is broad ($\sim 2000$ km s$^{-1}$) and
variable, the other is narrow ($\sim 100$ km s$^{-1}$) and constant. Both are produced by ionized and outflowing gas which obscures both the continuum and 
at least part of the Broad Emission Lines (BEL). 
Four years later the broad absorption line component was no longer 
detected in the spectra of NGC~3516 taken with IUE (from 1993, Koratkar et al. 1996), or later with HUT 
( in 1995, Kriss et al. 1996a) and GHRS (in 1995, Crenshaw, Moran 
\& Mushotzky 1998, Goad et al. 1999, hereinafter GEA99), suggesting either a change in the covering factor or in the 
ionization structure of the gas responsible for that feature, or an 
increase in the emissivity of the CIV BEL. 
The complex X-ray absorber was observed by ROSAT (in 1992, Mathur, Wilkes \&
Aldcroft 1997, hereinafter MWA97) and ASCA (Kriss et al. 1996b,
Reynolds 1997, George et al. 1998) through the detection of 
two strong absorption features at the energies of the OVII and OVIII 
K edges (0.74 and 0.87 keV respectively). The optical depths measured in 
the ROSAT and ASCA spectra using phenomenological 2-edge models, are similar 
to one another ($\tau_{OVII}\sim0.6$, $\tau_{OVIII}\sim0.3$).  

\medskip
Neutral absorption has not been unambiguously found in this source.
While the 1979 {\it Einstein} observation (Kruper, Urry \& Canizares 1990)
did not detect absorption by neutral gas, a subsequent low energy resolution
 EXOSAT observation in 1985 showed
the presence of cold gas, with N$_{H}\sim$10$^{22}$ cm$^{-2}$ affecting
the low energy spectrum (Ghosh \& Soundararajaperumal 1991).
A 1989 {\it Ginga} observation (simultaneous to IUE, Kolman et al 1993),
showed a low energy cutoff of the intrinsic power law, indicating 
heavy absorption of the nuclear continuum along the line of sight. 
However, due to the limited energy coverage of {\it Ginga} (2-20 keV 
bandpass) it was not possible to establish the ionization state of that 
absorber, which was consistent with either a high column of neutral gas 
($N_H^{Cold}$= 4$\times 10^{22}$ cm$^{-2}$) or a lower column of more 
highly ionized gas.

\bigskip
In this paper we present data from two broad band BeppoSAX (Boella et al. 1997a) observations of NGC~3516. 
Preliminary results were presented by Stirpe et al. (1998). 
The paper is organized as follows. In \S 2 and \S 3 we present our analysis 
of the data, while in \S 4 we discuss our main results. Finally, we 
present our conclusions in \S 5. 

\section{Data Reduction and Analysis}
NGC~3516 was detected with a good signal-to-noise ratio in the three main Narrow 
Field Instruments (NFI) on BeppoSAX, spanning 0.1-120 keV. 
In Table 1 dates and exposure times of the two BeppoSAX 
observations of NGC~3516 are presented, along with the source count rates measured in the LECS (Parmar et al. 1997), MECS (Boella et al. 1997b) and 
PDS (Frontera et al. 1997). 
Data from these three instruments were screened following standard 
criteria, as detailed in Fiore, Guainazzi \& Grandi (1999). In particular, 
PDS data were screened using fixed rise-time thresholds. 
Data from the three MECS units were merged to improve the statistics. 
The scientific products for the analysis are extracted from the equalized 
event files, using both the standard FTOOLS package (vs. 4.2) and the 
new release (vs. 1.1) of the CIAO (Chandra Interactive Analysis of 
Observation) software (Elvis et al. 2000). 
LECS and MECS source counts were extracted from circular 
regions of 6' radius, centered on the source. 
The same extraction regions were used to extract background counts from 
the long exposure ``blank-fields'' provided by the BeppoSAX Science 
Data Center\footnote{\tt http://www.sdc.asi.it}. 
PDS source and background counts were collected, respectively, during 
the on- and off-source modes of the instrument. 
For all the instruments, spectra were binned following the instrumental 
resolutions and allowing a maximum of 3 channels per resolution 
element.

\subsection{Model-Independent Evidence for Long-Term Spectral Variability}

NGC~3516 underwent moderate (factor of 1.5--2) 0.1--10 keV flux variations 
during both BeppoSAX observations (on a timescale of hours), 
however the ratios of the soft (0.1--2 keV) to the hard (2--10 keV) 
lightcurves were consistent with being constant, indicating no 
significant spectral variability across these energy bands on a 1-day 
timescale (the duration of each observation). 

A dramatically different behaviour was observed on the longer 4 months 
timescale between the two observations, during which the source experienced 
strong spectral variability. 
The 0.1--2 keV count rate increased by a factor of $\sim$4, while the 
corresponding increase in the 2-10 keV band was only of 1.9 (Table 1).
Figure 1 shows the model-independent ratio between the raw LECS 
(0.1--2 keV), MECS (1.5-10 keV) and PDS (13-100 keV) spectra for the 
two BeppoSAX observations (1996 November / 1997 March). 
The amplitude of non-linear effects in the responses of the NFIs 
(potentially affecting the results of such an analysis) is, at 
all energies, much smaller than the amplitude of the features 
visible in this ratio (Boella et al. 1997b, Parmar et al. 1997, 
Frontera et al. 1997). 

In the ratio in Figure 1, the 1--5 keV energy band is a smoothly 
increasing function of energy. It then flattens at E$\gs 5$ keV, 
around a value of $\sim 0.6$. In the PDS band the ratio is $\sim$0.8. This implies a steepening of the ratio 
between 10 and 13 keV. 
A sharp emission feature around 1 keV is visible superimposed on these smooth changes. 
All this strongly suggests that some, and perhaps all, of the several 
components that make up the 0.1-100 keV spectrum of NGC~3516 varied independently of each other from one observation to the other. 
In particular, the smooth and monotonic rising of the ratio up to 
$\sim 5$ keV, along with its flattening above this energy and up to 
10 keV, suggest a drastic decrease in the degree of 
obscuration of the central X-ray source between 
the two observations rather than an intrinsic 
variation of the slope of the primary X-ray continuum. In this case the asymptotic value of $\sim 0.6$
around which the ratio flattens above $\sim 5$ keV, 
would be the intensity increase 
factor of the intrinsic X-ray power law between the two observations. 
The higher PDS count ratio of $\sim$0.8 may suggest a higher relative
intensity of a Compton-reflection component in the 1997 observation.
There is no clear indication for a similar effect at the energy of the 
Fe K line. 
The emission feature at $\sim 1$ keV may be the signature of iron L 
emission, visible only during the 1996 observation, and therefore either external to the nuclear
environment, and detectable by the BeppoSAX
LECS only when the soft nuclear continuum was heavily obscured, or 
internal to that, but only partially obscured by the amount of neutral 
(or mildly ionized) gas obscuring the X-ray source in 1996. Intrinsic 
variation of the intensity of the Fe-L complex in the hot plasma is, of 
course, a further possibility. 

\noindent
We use this scenario to guide the modelling and fitting of the spectra 
in the following sections. 

\section{Spectral Modelling and Fitting}
We have performed global modeling and fitting of the 0.1-150 keV 
spectra of both BeppoSAX observations of NGC~3516 using the 
XSPEC 10.0 package. In Table 2 we summarize the results of this 
analysis. Errors are quoted at 90\% confidence level, for one 
interesting parameter (i.e. $\Delta\chi^2 = 2.71$, Avni 1976). The
cosmology used is H$_{0}$=50 km s$^{-1}$ Mpc$^{-1}$ and q$_{0}$=0.5. 
For all the absorption (neutral and/or ionized) models that we use 
in this paper we adopt solar abundances as estimated by Grevesse and  
Anders (1989) and Grevesse and Noels (1993). 
Photoionization models for fitting purposes include 
photoelectric absorption but not resonant absorption and gas emission, 
and have been built following Nicastro et al. (2000). 
Finally, we use pure Compton reflection models built by running iteratively 
the routine {\tt pexrav} (Magdziarz \& Zdziarski 1995) for the inclination angle of 30$^{o}$
(consistent with the best fit interval measured, see \S 3.1.3) and a range of values of the slope $\Gamma$ of the
incident continuum and the e-folding energy $E_c$ of the high energy 
exponential cut-off. 
For fitting purposes we always combine the Compton reflection model with 
a cut-off power law, linking the two $\Gamma$ and $E_c$ to the same 
values, and leaving the two normalizations free to vary independently. 

\bigskip
We first fitted the 1996 November and 1997 March spectra 
(hereinafter N96 and M97) with a model consisting of a power law 
plus Galactic absorption (N$_{H}$= 3.4$\times 10^{20}$ cm$^{-2}$, 
Dickey \& Lockman 1990). 
Both fits are very poor ($\chi^2/dof (N96) = 1147/94, 
\chi^2/dof (M97) = 600/101$), and the best fit photon indices are 
different: $\Gamma \sim 1.2$ and $1.5$, respectively. 
The residuals clearly show the presence of several additional 
non-modelled spectral features (Figure 2a and b). 
Some are common to both spectra: (a) a line-like 
feature at $\sim 6.4$ keV, (b) a deficit of counts at $\sim 
8$ keV, and (c) a systematic deviation across the PDS band. 
However, at low energy, (E$\ls 4$ keV) the shapes of 
the residuals are quite different from one spectrum to the other. 
Our best fitting parameterizations of N96 and M97  
are thus quite different, and are summarized in Table 2. 

The two Best Fitting models (hereinafter 96BF and 97BF respectively) 
include several common components, which make up our ``Base Model'': 
(a) a power law with a high energy cut-off; (b) a neutral absorber 
(to account for Galactic absorption along the line of sight); 
(c) a hot photoionized absorber (HA, see Table 2), parameterized by a
column density $N_H$ and an ionization parameter U (to account for the
8 keV deficit); (d) a Gaussian emission line (to account for K$\alpha$ 
iron fluorescence); (e) a pure Compton-reflection component (limited to 
$\Gamma$ and $E_c$, as noted earlier).  
However, N96 and M97 differ from each other in requiring 
additional components that are not held in common: (f) a neutral 
absorber only partially obscuring the direct view of the 
primary X-ray power law and (g) soft emission by an optically thin 
plasma, for the N96 spectrum, and (h) a second, more mildly ionized 
absorber, for the M97 spectrum. 

\subsection{The ``Base Model''}
Our estimates for the parameter values of our ``Base Model'' were 
consistent with each other between the 96BF and the 97BF 
parameterizations (see Table 2). 
In the following we describe and discuss 
the individual components of this model referring to the numerical 
results obtained for the M97 spectrum, the one with higher statistics. 

\subsubsection{The Intrinsic Power Law and the High Energy Cut-Off}
The 0.1--100 keV primary continuum of NGC~3516 is well 
described by a simple power law $F(E) = AE^{-\Gamma}$ (F(E) in ph 
s$^{-1}$ cm$^{-2}$ keV$^{-1}$) with $\Gamma = 
2.04^{+0.06}_{-0.07}$ (Table 2). 
This component varied in flux by $\sim 70\%$ between the 
two observations (i.e. A(N96)/A(M97) = 0.6), without changing in shape. 
We then conclude that little, if any, of the dramatic spectral 
variability experienced by the source can be due to variation of 
the primary continuum, either in shape or intensity. 

\medskip 
We could only estimate a lower limit (E$> 450$ keV) on the $e$-folding 
energy of the high energy cut-off, and were not able to measure 
any significant change of this parameter, between the two observations. 

\subsubsection{Iron Line and Compton Reflection}
Both the iron line and the Compton-reflection components were required 
by our data at high significance. Removing each of these components, 
one at time, from the 97BF model and refitting the data gives, 
respectively (with $\Delta\nu$ indicating the number of free parameters 
eliminated by the 97BF model): $\Delta\chi^2 = - 25$ ($\Delta\nu = 3$) 
and $\Delta\chi^2 = -41$ ($\Delta\nu = 1$). 

\medskip
The best fit energy of the iron emission line (E = 
$6.41\pm 0.15$ keV) is consistent with that of 
FeI-XXII K$\alpha$ transitions. 
When modelled with a simple Gaussian (as in our adopted best fitting 
parameterization 97BF), the line is consistent with being narrow, with 
a 90\% upper limit on its width of 0.27 keV. 
We note that the energy resolution of the 
BeppoSAX-MECS is only of $\sim 500$ eV at 6.4 keV. 
Furthermore, an accurate determination of the exact 
shape of the iron line profile is hampered by 
the presence in the data of the $\sim 8$ keV edge-like absorption 
feature. 
However, we have checked the consistency of our data with 
more complex scenarios, replacing the Gaussian emission line in 
the 97BF model, with a relativistically broadened and distorted 
emission line from the accretion disk (the model {\tt diskline} in XSPEC: 
Fabian et al. 1989), as suggested by previous ASCA observations (Nandra 
et al. 1997a,1999). 
We fixed the internal radius on the accretion disk 
at 6 gravitational radii, the energy of the emission line centroid 
at 6.33 keV (the best fit energy in 97BF) and the spectral index of 
the emissivity law at -2, and refitted the data. 
The result was inconclusive. We obtained the same $\chi^2/\nu = 
86/95$ as in 97BF. The outer radius on the disk was only 
barely constrained and larger than 11.5 gravitational radii. 
All the other model parameters were consistent within the errors 
with the 97BF measurements. 
We also checked the extreme hypothesis that all of the negative 
residuals visible in Figure 2 at $\sim 8$ keV could be explained 
in terms of a very distorted blue-profile of the iron line, 
eliminating the HA component and refitting the data. This yielded 
a worse $\chi^2$ than 97BF ($\Delta\chi^2 = - 18$, for $\Delta\nu = 3$), 
and accounted only partially for the absorption feature at $\sim 8$ keV. 
We then conclude that a relativistically broadened and 
distorted emission line profile with no FeXVII-XXII K absorption 
is not a necessary component for the description of our data. 

\bigskip
Thanks to the broad band of BeppoSAX we were able to detect 
a bump in the 10-30 keV data (see Figure 2a and 2b). 
We model this bump as due to Compton reflection of the 
primary power law off a slab of almost neutral matter re-radiating the 
isotropic incident X-ray continuum. 
During both the 1996 and 1997 observations we measured relative 
fractions of reflection consistent with each other (Table 2), 
and fully consistent with the value of 1 corresponding to a solid 
angle of 0.5 as seen by a central isotropic source of primary radiation. 
Their ratio, $0.8_{-0.2}^{+0.7}$, is consistent with the value 
derived in \S 2.1 from our model independent analysis. 
If the same matter were also producing the K$\alpha$ fluorescence iron 
line (but see also \S 2.1), we would expect the equivalent width of 
this line to be $\sim$100-150 eV (e.g. Mushostzky, Done \&
Pounds 1993), consistent with the measured value (see Table 2). 

\subsubsection{The Hot Absorber}
Both the N96 and M97 spectra of NGC~3516 
show a clear absorption feature around 8 keV (Figure 2). A neutral reflection component alone cannot explain the edge feature at $\sim$8 keV, even
allowing the inclination disk parameter to vary in a wide range (from 0$^{o}$ to 60$^{o}$).
We accounted for this feature by including a hot photoionized absorber 
(HA) in 96BF and 97BF, obscuring the line of sight. The same model was used for NGC~3516
to fit the absorption feature detected at the same energy (E$\sim$8 keV) by {\it Ginga} (Kolman
et al. 1993).
We stress again that the model used to fit the data 
includes neither gas emission, nor resonant absorption. 
Removing this component from both 96BF and 97BF and refitting 
the data, gives respectively $\Delta\chi^2 = - 13$ and $\Delta\chi^2 
= - 29$, for $\Delta\nu = 2$. 

The best fitting parameters U and $N_H$ do not vary (within the 
errors) between the two observations (Table 2), and their values 
indicate that a high column density ($N_H^{HA} = (3.2_{-0.6}^{+0.5}) 
\times 10^{23}$ cm$^{-2}$, very similar to the value found by Kolman
$\sim 3\times 10^{23}$ cm$^{-2}$) of highly ionized (U = 224$_{-83}^{+131}$)
gas is absorbing the primary continuum of NGC~3516 in both data sets. 
In such an absorber elements lighter than oxygen and neon are fully 
ionized, and therefore do not imprint any visible features on the emerging 
spectrum. 
However, iron is still not fully ionized and contributes strongly 
to the opacity of the absorber at the energies of the photoelectric 
K and L absorption edges of FeXVII-XXII. 
Absorption by FeXVII-XXII K imprints deep absorption edges at 
$\sim 7.8-8$ keV on the emerging spectrum, while L absorption 
by the same species affects the whole 1 to 5 keV band, gently 
flattening a low-resolution spectrum rather than producing visible 
sharp and resolved single absorption features on it (unless Fe-L 
resonant absorption lines dominate on recombination lines
\footnote{The relative contribution of Fe-L resonant absorption lines, 
compared to emission lines, depends critically on the geometry and 
dynamics: if the gas has a high microturbulence velocity and a lower or 
comparable bulk-velocity, then Fe-L resonant absorption dominates 
and imprints a visible ``1-keV'' feature on low resolution spectra 
(Nicastro, Fiore \& Matt, 1999). See also \S 4.2}). Fitting low resolution data with a phenomenological single-edge 
model, accounting for the $\sim 8$ keV absorption feature, would 
result in a wrong estimate of the slope of the intrinsic 0.1-100 keV 
power law, which would appear to be flatter by $\Delta\Gamma = 0.28 \pm 0.13$. Ionized, and possibly relativistically smeared, edge-like features at energies $>$7.1 keV
have often been detected in the hard/low state of Galactic Black Hole Candidates, where the
accretion disk is supposed to be hotter than that of AGNs by a factor $\sim$30 (Done \& $\dot{Z}$ycki
1999, Ross et al. 1996). In this contest, to further test the uniqueness of our interpretation,
we applied an ionized reflection model, which possibly can also account for an ionized
edge-feature, to our data.
The $\Delta\chi^{2}$ between a model with an ionized reflection (model {\tt pexriv} in XSPEC,
Magdziarz \& Zdziarski 1995) and the best fit (BF)
is $\Delta\chi^{2}/\Delta\nu$=25/1 and $\Delta\chi^{2}/\Delta\nu$=23/2 if the iron abundance
parameter is left free to vary. The variation of the inclination angle of the ionized reflector
($\theta= 30^{o}, 45^{o}, 60^{o}$) does not influence the $\chi^{2}$ test. Moreover, the expected
ionized iron line at 6.7 keV, consistent with the ionization degree of the edge, is not detected,
even though a ionized line could have been missed due to the low resolution instrument. 

The preferred interpretation is thus in terms of a high ionization, high column density absorber
which can simultaneously account for both the $\sim$8 keV edge and the overall continuum shape at
medium energies and, most important, it provides the statistically better description of the available
 data.

\subsection{Departure from the ``Base Model''} 
As pointed out in \S 3 the ``Base Model'' does not provide a fully satisfactory description of either the N96 or the M97 spectra. 
Additional components are present in our best fitting 
parametrizations 96BF and 97BF. We describe them in the following 
two sections. 

\subsubsection{The ``Cold'' Absorber and the Soft Thermal Emission in
N96}
To model the 0.1-5 keV residuals of Figure 2b, we added to our ``Base 
Model'' a neutral absorber of column density $N_H^{Cold}$ which only 
covers a fraction C$_{sou}$ of the nuclear X-ray source. 
Fitting the N96 data with this model produces a significative 
improvement in the $\chi^2$ ($\Delta\chi^2/\Delta\nu = 73/2$) and gives 
N$_H^{Cold} = 2.31^{+0.33}_{-0.15} \times 10^{22}$ cm$^{-2}$ and $0.7<C_{sou}<1$. (We note that using an ionized absorber model the ionization parameter is consistent to be zero, while the column density of the warm gas is $\sim N_H^{Cold}$, indicating absorption by neutral gas as the most likely explanation). Nevertheless, the residuals continue to show a clear line-like unresolved 
feature around 1 keV (Figure 3, upper panel). 

\noindent
At this energy, recombination L lines from FeXVII-XXIV are expected 
to strongly contribute to the emission from either a collisionally 
ionized plasma (CIP) with a temperature of $\sim 1-2$ keV, or from a 
photoionized plasma (PIP) with a high ionization degree (U$\sim 200$) 
and column density ($N_H \sim 10^{23} cm^{-2}$). 
Only for fitting purposes, we use here a CIP model, 
while a more accurate discussion on the physics of this component and 
on its possible identification is deferred to \S4.2. 
We added a CIP component (the model {\tt mekal} in XSPEC, Mewe,
Gronenschild \& van den Oord 1985) to our 
model, and refitted the data. 
The fit is again significatively improved by the addition of this 
component ($\Delta\chi^2 = 13$, for 2 additional parameters: model 96BF), 
and the residuals are now flat over the entire 0.1-100 keV band. 
The best fit temperature of this collisional plasma 
is $kT = 1.78^{+0.91}_{-0.39}$ keV. The covering factor of the 
neutral gas is now larger ($C_{sou} > 0.95$) than in the previous fit, 
indicating that only a small or null fraction of the nuclear radiation escapes absorption.

\subsubsection{The ``Warm'' Absorber in M97}

Our ``Base Model'' did not provide a satisfactory description 
for the M97 spectrum either. 
Several features remained at E$\ls 2$ keV as shown by the residuals 
in Figure 3b (bottom panel). These strongly suggest the presence of 
a typical ``warm'' absorber obscuring the line of sight and imprinting 
deep OVII and OVIII K edges, at 0.74 and 0.87 keV (see Reynolds 1997, 
George et al. 1998). For this source the signature of a complex ``warm'' absorber had been 
previously detected by ROSAT-PSPC and ASCA (MWA97, 
Kriss et al., 1996b). 
We then added a photoionized absorber component to our ``Base Model'' 
and refitted the M97 data. The residuals are now flat over the 
entire BeppoSAX band, and the reduced $\chi^2$ is 0.90 (Table 2: 
97BF model). 

\section{Discussion}

\subsection{The Variable Absorber}
The 1996 November spectrum of NGC~3516 is heavily 
absorbed by almost neutral gas which covers most of the primary 
continuum source, (here we choose C$_{sou}\sim 0.95$, the best fit value obtained fitting
the soft emission with the CIP component, \S 3.2.1),
and whose equivalent hydrogen column 
density (as derived by a fit with a neutral absorber) is as large 
as $N_H^{Cold} = 2.3 \times 10^{22} cm^{-2}$ (exceeding the Galactic 
value by almost two orders of magnitude). This value would be even 
larger if the gas were mildly ionized (OII-III dominating). 
There is no evidence for variation of the degree of obscuration 
of this gas during the observation ($\sim$20 hours) 
and  this component is not required by the 1997 March data (4 months 
later)
\footnote{We note, however, that an ionized, similar $N_H$, absorber 
is needed in 1997: see \S 4.1.2.}, 
and the UV continuum of this source in a HST/FOS observation taken 
$\sim 18$ day after the 1996 November BeppoSAX observation (GEA99) shows no marked extinction by dust (assuming a Galactic 
gas/dust ratio we should expect $A_v \sim 10$ mag, for $N_H = 2.3\times 
10^{22} cm^{-2}$, Gorenstein, 1975). 

All this allows us to put constraints on the geometry and 
the physical state of this component, and to speculate on 
different possible scenarios. 

\subsubsection{Transient Cold Absorption by a Broad Emission Line Cloud?}

The typical timescale of variability of the X-ray continuum (e.g. Nandra et al. 1997b) gives 
an upper limit on the linear size of the emitting region: $D \sim 
c \Delta t \sim 10^{14}$ cm. 
In order for an homogeneous cloud of 
absorbing gas to cover almost the entire primary emitting region, 
either the cloud is at a distance $R$ close to the central source, comparable to $D$, and its linear sizes $l$ are comparable and/or 
larger than $D$, or $R$ is much larger than $D$ and the source 
appears point-like as seen by the cloud.  
Photoionization arguments strongly support the second picture. 
In fact, assuming $R\sim D$  gives an ionization parameter U$>$ 2500 n$_{10}^{-1}$, where 
n$_{10}$ is the electron density of the gas 
in units of 10$^{10}$ cm$^{-3}$. We used the ionizing luminosity of 
NGC~3516, derived assuming the 2200 \AA\ flux and $\alpha_{ox}$ 
from George et al. (1998) and X-ray fluxes and spectral shape 
from our data. 
Such an ionization parameter 
would be associated to almost neutral gas (U$\ls 0.01$: oxygen 
distributed between OI-OIII) only for implausibly high electron 
densities
\footnote{We note that this is one of the highest 
densities one may expect to find in the nuclear environment of a Seyfert 
1, being of the order of the density of the matter in a standard 
Shakura-Sunyaev (1973) accretion disk around a black hole of $10^7-10^8$ 
M$_{\odot}$, accreting at 0.1-0.01 times the critical rate, and in 
proximity of the last stable orbit.} 
$n_{10} > 2.5 \times 10^{5}$. 
More plausible electron densities of $n_{10} = 0.1-100$, as those 
estimated for the Broad Emission Line Clouds (BELCs) in AGNs (Netzer
1991, Ferland et al. 1992, Krolik et al. 1991), would result in very highly ionized 
gas, which is inconsistent with our data. We then assume that $R >>D$. 

Reverberation mapping studies of the BELs in NGC~3516 (Wanders et al. 1993) have estimated a distance $R_{BELC}$ of the BELCs 
from the source of ionizing continuum, of $\sim 11$ {\em ld} ($\sim 3 
\times 10^{16}$ cm), supporting 
the possibility that the gas obscuring the X-ray continuum of 
NGC~3516 in 1996 is associated with a BELC. 
At this distance, and assuming $n_{10}^{BELC} = 0.1-100$, 
the ionization parameter measures U$(R_{BELC}) \sim 0.0003-0.03$, 
and the gas is only mildly ionized and therefore compatible with the 
nature of the absorption seen in the X-ray band in 1996. 

All this suggests that a BELC crossing our line of sight could have 
caused the obscuration of the X-ray primary source in 1996. 
As a consistency check we can calculate the time $t_{cross}$ 
needed for a BELC to cross our line of sight. 
Assuming as transverse velocity $v_{BELC}$, the measured H$\beta$ 
FWHM of 4,000 km s$^{-1}$ (Wanders, et al, 1993), we obtain $t_{cross} = l/v \simeq 2,500 N_{H^{Cold}_{22}} n_{10}^{-1} 
v_{4000}^{-1}$ (where $l$ is the linear size of the cloud). 
For the measured N$_H^{Cold}$, we have then $t_{cross} \sim 6,000 
n_{10}^{-1}$ s. Since we know that $t_{cross}$ must 
be longer than 20 hours (the duration of the 1996 observation) and 
shorter than $\sim 4$ months (the time elapsed between the 
two BeppoSAX observations), the above equation constrains the electron density of the gas to the range 
$6 \times 10^{-4} \hbox{ cm}^{-3} \ls n_{10} \ls 0.1$, 
We stress again that, if the gas were mildly ionized (as is the 
case for the BELCs), then we would be underestimating the equivalent 
N$^{Cold}_H$ obtained fitting the data with a neutral absorber model 
and therefore the interval of densities given above would shift towards higher 
values. This range of densities contains the estimated BELC 
value. 
A much narrower interval could in principle be set using the time 
elapsed between the 1996 November observation and the last HST/FOS 
observation of GEA99), showing no MgII resonant absorption 
nor extinction in the UV continuum, in $\Delta t = 18$ days. 
This would give a lower limit on the density of $n_{10} \gs 0.01$. 

\medskip
In the framework of the proposed scenario, a BELC crossing the 
line of sight, we estimated the probability of such an 
event, based on an estimate of the filling factor of the 
BELCs. The total volume of a sphere with radius $R_{BELC}$ 
is $V_{tot}\sim 8.2\times 10^{49} cm^{3}$. 
From the luminosity of the broad H$\beta$ of NGC~3516, L$_{H\beta}\sim
2.3\times 10^{40} erg\ s^{-1}$ (Rosenblatt et al. 1994), we can 
estimate the effective volume occupied by all the BELCs, given by:\\
$V_{eff}=L_{H\beta}/\alpha(T_e)n_e^2$. For an electronic temperature
T$_e$ of $\sim 10^{4}$ K, at the energy of the H$\beta$ emission
line, the recombination coefficent $\alpha(T_e)$ is $\sim$3.4$\times
10^{-25}$ erg cm$^{3}$ s$^{-1}$ (Osterbrock 1989) and $n_e\sim 10^{10} cm^{-3}$ (Netzer 1991). 
We then obtain $V_{eff}\sim 6.7\times 10^{44} cm^{3}$. The probability for 
a cloud to find itself along a particular line of sight at any radius 
is then given by $P \sim V_{eff} / V_{tot} \sim 8.1\times 10^{-6}$. 
This low probability supports the picture of an event which occurs 
very rarely, as indeed observed in AGN (Kruper, Urry \& Canizares 1990). 

\subsubsection{A Variable State Absorber?}
Another intriguing possibility is that the ionization state 
of the absorber varied between the two BeppoSAX 
observations, becoming more ionized, and therefore transparent to the soft 
X-ray radiation.  
In this scenario we may speculate that the ``warm'' absorber seen 
in the 1997 March spectrum of NGC~3516 is the same gas which was 
obscuring the line of sight 4 months earlier, but with a very different 
ionization degree. The same picture was drawn for this source 
by MWA97 who proposed a model for a variable ``warm'' 
absorber in NGC~3516 to reconcile the UV and the X-ray spectra within 
a unique scenario (see also next section). 
In the MWA97 model the presence and subsequent 
disappearance of a broad CIV absorption component at 1549 \AA 
was consistent with a variable state X-ray absorber, from mildly 
ionized during the 1989 {\it Ginga} observation to highly ionized in 
subsequent ROSAT-PSPC and ASCA observations. 

In the framework of the scenario above described, we derived some dynamical properties of this variable state absorber. 
At a given ionizing flux, the ionization degree of a photoionized 
absorber can vary due to variations of the distance of the gas from 
the ionizing source, or of the electron density in the gas (or both). 
In the first of these two extreme hypotheses we assume that 
the absorber has a constant density in the range $n_{10} = 10^{-4} 
-10^{-1}$ (typical of ``warm'' absorbers in Seyfert 1s: Nicastro 
et al. 1999, 2000, MWA97). 
The gas is almost neutral during the 1996 November observation, 
which gives an upper limit on its ionization parameter of $U \ls 
0.01$, above this value, and for the ionizing continuum shape of 
NGC~3516, the opacity of the gas at low X-ray energies drops 
drastically. This gives a range of possible distances from the 
central source of $R \gs 2 \times 10^{17} - 5 \times 10^{18}$ 
cm (the larger the density the smaller the lower limit on $R$). 
If the gas were moving toward the central source with a radial 
velocity $v_{rad}$, keeping its density constant, it should 
have required less than $t \ls 4$ months to cover the distance $\Delta R = 
R/10$ needed for its ionization parameter to increase by a factor of 
$\sim 100$ (to reach the values observed during the 1997 
November observation). The estimated 
radial velocity of the gas is thus $v_{rad} \sim \Delta R / t 
\gs 0.06 c$. This limit would become even higher if the cloud became 
denser while collapsing toward the central source. We note that the most direct implication of this extreme hypothesis is that UV and high
resolution X-ray spectra of Sy~1 galaxies, during such relatively shorts events, should show highly
redshifted resonant absorption lines with varying strenght. Such phenomenon has never been observed
so far.

\medskip
Alternatively, the change in the ionization state of the absorber
might be a result of a change in density (as in MWA97). To
change from an almost neutral state (U\lax 0.01) to a warm state
($U\sim5.37$) would require a drop in density 
(U$\propto n_H^{-1}$) by a factor of \gax 500. For example, the density of a BELC would have 
to change from a few 10$^{10}$ cm$^{-3}$ to about $2\times10^{7}$ 
cm$^{-3}$. A column density of $10^{22}$ 
cm$^{-2}$ and a density of 10$^{10}$ cm$^{-3}$ imply a thickness 
of $10^{12}$ cm for the cold absorber. Similarly, the thickness of the warm cloud would be about
$5\times 10^{14}$ cm. Such an expansion has to occur on a time
scale of \lax 4 months. If the cloud is expanding adiabatically, its
velocity dispersion would then have to be about 500 km s$^{-1}$. Such
a velocity dispersion is observed in associated absorption lines in AGN,
so the derived value is completely reasonable. We conclude that a 
scenario based on a variable state absorber is plausible (see \S 4.3 for 
further details). Unfortunately, no other X-ray observation was performed between the two BeppoSAX pointings to further constrain the variability time scale in the soft X-ray energy band.

Both cases discussed above were based on the assumption of no 
changes of the intensity of the photoionizing continuum. 
Variations of the ionizing flux could have contributed to 
the observed changes of the ionization degree of the gas. 
If, between the two observations, the continuum increased by a factor 
of $\sim 100$ and then recovered the level measured during the 
1997 observation, then the gas could have been almost instantaneously 
ionized by this dramatic event, but not yet had enough time to 
recombine to its equilibrium neutral state. Alternatively, the flux 
level seen during the 1996 observation could have been preceeded by a 
long duration event of very low intensity which caused the gas 
to recombine to an almost neutral state. The gas during the 1996 
observation would then be underionized with respect to the observed 
ionizing continuum level. Both cases would require low electron densities 
of the absorber ($n_e \ls 10^5$ cm$^{-3}$, see Nicastro et al. 1999, 
for details). 

\subsection{A hot Ionized Absorber/Emitter gas?}\label{high:U}
Two apparently unrelated spectral features are clearly visible in 
the N96 spectrum of NGC~3516: (a) soft emission (with a hump at 
$\sim 1$ keV), not obscured (or only partly obscured) by the amount 
of neutral gas covering the primary X-ray source, and (b) the 
signature of heavy absorption by highly ionized gas at $\sim 8$ 
keV. In our spectral analysis of N96 we modeled these two 
features as due to two distinct components: a collisionally 
ionized emitting plasma and a photoionized absorber, respectively. 
Extended emission by the host galaxy surrounding the AGN has 
frequently been observed in Seyferts: e.g. Elvis et al. (1990), 
Wilson et al. (1992), Weaver et al. (1995).
However, in the particular case of NGC~3516, the extended component 
was probably unresolved by the ROSAT-HRI (Morse et al. 1995). 
Wilson et al. (1996) found that in their sample (which includes 
NGC~3516) the 0.1--2.5 keV intrinsic luminosity of the extended components 
ranges between $10^{40}$ and $10^{42} erg\ s^{-1}$. In the present data we measure 
$L_{0.1-2.5 keV} = 5 \times 10^{42}$ erg s$^{-1}$, a value 
only marginally consistent with the observed range, even allowing 
for intrinsic variability. 
Based on the correlations found between far infrared (FIR) and 2 keV 
luminosity of bright spiral galaxies (Fabbiano et al. 1988), 
we also compared the 2 keV luminosity $L_{2 keV} = 9 \times 
10^{41}$ erg s$^{-1}$ keV$^{-1}$ measured, with the mean $L_{2 keV}$ value 
for the 11 objects of the Fabbiano et al. (1988) sample with 
$L_{FIR} = 10^{42.8}-10^{43.2}$ erg $s^{-1}$. This range includes 
the value of $10^{43}$ erg $s^{-1}$ obtained for NGC~3516 
(Bonatto \& Pastoriza 1997). 
We found $L_{2 keV}^{Gal} \simeq 10^{39} - 10^{40} erg\ s^{-1}$, 
again much lower than measured in our data. 
We then conclude that the soft X-ray emission visible 
in 1996, cannot be entirely due to an extended component 
in the host galaxy. 

An intriguing alternative is that both the 
$\sim 8$ keV absorption edges visible in N96 and M97 and the soft 
emission in N96 could be self-consistently accounted for by a 
unique photoionized absorber/emitter in the nuclear environment of 
NGC~3516. 
If photoionized, this gas has a temperature of $T\simeq 2 \times 
10^6$ K, and iron (the lightest abundant element not yet fully ionized) 
is smoothly distributed between FeXVII and FeXXVI. Due to the high 
temperature the emissivity of this gas is large and contributes to 
the overall spectrum in an amount which depends linearly on its 
covering factor around the central source. 
In particular, a strong contribution by FeXVII-XXIV L recombination 
lines is expected at $\sim 1$ keV, where a clear emission feature is 
present in the N96 spectrum. 
Models for photoionized gas accounting for gas emission, other than 
photoelectric and resonant absorption, can then be used to constrain 
the geometrical configuration of the observed absorber/emitter. 

In the following we try to test this hypothesis quantitatively. 
The basic assumptions of this hypothesis are: (a) 
the neutral absorber in 1996 covered almost entirely 
($C_{sou}> 95$ \%, as from the fit with the CIP component, \S 3.2.1), the nuclear source
of primary X-rays
but left uncovered a much larger fraction $(1 - C_{env})$ 
of the extended nuclear environment (i.e. the warm and hot scattering
 media) which is seen through scattering of the primary continuum; 
(b) during the 1997 observation both the primary nuclear continuum and 
the reprocessed components are seen directly (c) no other component
can significantly contribute to the line emission.

\medskip
We proceeded to build (following Nicastro et al. 2000) two series of 
photoionization models, which include gas emission and resonant 
absorption, for the two ionized absorbers observed in NGC~3516. We used 
column densities and ionization parameters as measured fitting the 
N96 and M97 spectra with models accounting for photoelectric absorption 
only (Table 2). We built these models for a number of values of the 
covering factors $f_{warm}$ and $f_{hot}$ as seen by the central 
source, varying from 0 to 1, and for particular values of outflowing ($v_{out}$) 
and microturbulence ($\sigma$) velocities of the absorbers/emitters. 
In particular we used $v_{warm}^{out} = 500$ km s$^{-1}$, 
$\sigma_{warm} = 1,100$ km s$^{-1}$ (observed in the UV for the 
broad CIV system of NGC~3516, Kolman et al. 1993), and $v_{hot}^{out} 
= 500$ km s$^{-1}$,$\sigma_{hot} = 200$ km s$^{-1}$. 
For the ``hot'' absorber the adopted ratio $v_{hot}^{out}/\sigma_{hot}$ 
maximizes the contribution of line emission compared to resonant absorption 
(Nicastro et al. 2000). The particular choice of the absolute 
values of $v$ and $\sigma$, separately, is however not critical. 
The warm component was not included in the models 
for the N96 spectrum, since we did not see it in the data (which can 
be either because the component was not present or because it was completely 
obscured by the amount of almost neutral gas covering the line of sight 
during that observation). 
We also neglected the emission contribution from this gas. However 
given the temperature and ionization degree measured in 1997 
(in the photoionization hypothesis), emission of the warm component 
is expected to be very weak, and should influence mostly the 0.5-0.6 
keV portion of the spectra where weak OVII and OVIII emission 
lines are expected. This contribution would be hardly detectable 
with BeppoSAX (Nicastro et al. 2000). 
The hot component has been instead included in both models. 
We held $f_{hot}$ constant between the two observations. For the 
``cold'' absorber in 1996, we used the best fit column density and 
$C_{sou}$ from Table 2. 

We did not fit these models directly to the N96 and M97 spectra, but 
proceeded as follows. We varied the values of the three parameters 
$f_{hot}$, $f_{warm}$ and $C_{env}$ to let (a) the 0.1-2 keV flux 
predicted by models for N96 match the best fitting 96BF 0.1-2 keV flux, 
and (b) the 0.1-10 keV ratio between N96 and M97 models match the observed 
raw data ratio N96/M97. In this particular scenario, we could put relatively strong constraints on the covering factors
$f_{hot}$ and $C_{env}$, but not on $f_{warm}$, which we then fixed 
to the value of 0.5. We found that $f_{hot}$ and $C_{env}$ 
lie in the ranges 0.3-0.4 and 0.15-0.25 respectively. 
Figure 4 and 5 show two models for the N96 and M97 spectra of 
NGC~3516 (Fig. 4a,b, upper and bottom panels respectively), and 
their ratio (Fig. 5) in the energy band 0.1-10 keV, for 
$f_{hot}=0.35$ and $C_{env} = 0.2$. 
The ratio between the raw 0.1-10 keV N96 and M97 data is also 
plotted in Figure 5c. The agreement between the models and the data 
is quite good. In particular we note that the 1 keV emission feature 
present in the data is well reproduced by Fe L emission by the 
hot photoionized components, which was much more visible in 1996 due to the heavy obscuration of the primary X-rays by the 
``cold'' absorber. 

We then conclude that a highly ionized component is present in
the nuclear environment of NGC~3516, and that its emission is visible only when our view of the primary X-rays is
at least partially covered by a large amount of neutral absorber. 

This component could well be present in all Seyferts, but has eluded 
detection so far. 
If this component is spherically distributed around the central 
source, and if the covering factor $f_{hot}$ and column density 
$N_H^{hot}$ measured in NGC~3516 are representative of the entire 
class, then we would expect that absorption K edges due to highly 
ionized iron should be observed in almost 30-40\% of all known 
Seyferts. Furthermore, in all heavily obscured sources (i.e. Seyfert 
2s, or objects with transient absorption, like NGC~3516), we should 
detect emission by this hot gas, at energies below 1-2 keV. 
Soft X-ray emission is indeed frequently observed in 
many bright and well studied Seyfert 2s (Turner et al. 1997, Comastri 
et al. 1998). High resolution spectrometry, like that available with Chandra and/or 
XMM, will test this speculation unambiguously by allowing the 
detection of narrow emission and/or absorption lines even in 
objects with a low degree of nuclear obscuration with high accuracy, as already studied in NGC~5548 and NGC~3783 (Kaastra et al. 2000, Kaspi et al. 2000).    

\subsection{The Complex X/UV Connection}

Following MWA97, we investigate whether the X-ray 
absorbers in NGC3516 will produce any associated UV absorption
lines.

If the cold absorber is mildly ionized, the maximum degree of
ionization would be $U\sim 0.01$; as seen in \S4.1.2, a higher value, with  the ionizing continuum shape of 
NGC~3516, would imply a drastic drop of opacity of the gas at low X-ray energies. In such an absorber, the dominant ions
would be OI-III, MgII-III and CII-IV. For a total column density of a
few times $10^{22}$ cm$^{-2}$, the predicted column densities of MgII
and CIV are $2.5\times 10^{17}$ and $4.7\times 10^{17}$ cm$^{-2}$
respectively. For the warm absorber, the CIV column density would be
$5.4\times 10^{14}$, but Mg is too highly ionized to have observable 
MgII. For the cold absorber, the MgII and CIV column densities are
very large and even for a velocity dispersion parameter (b) of 100 km
s$^{-1}$, the resonance absorption lines would have large equivalent
widths (EW \gax 3\AA).

In the warm phase, the CIV column density is less than that in the 
cold absorber. However, the predicted EW
of the CIV $\lambda 1549$ absorption line would still be detectable
(EW\gax 1.0 \AA, ~for b=100 km s$^{-1}$). The absorption line EWs
would be even larger for larger b. Therefore, for reasonable 
parameters, the X-ray absorbers in NGC~3516 will contribute to the
absorption in UV. How broad the lines would be will again depend upon
the b parameter. The UV observations closest in time to our X-ray 
observations ($\Delta t = 18$ days) are presented by GEA99 ($\S 4.1$). 
With the HST FOS, intrinsic MgII absorption was not detected. 
This constrains the cold X-ray absorber (if still along the line of 
sight during the UV observation) to be almost completely neutral. 
GEA99 found CIV $\lambda 1549, 1551$ absorption doublets with
FWHM 651$\pm$16 km s$^{-1}$ and 405$\pm$9 km s$^{-1}$ respectively and EWs
2.87$\pm$0.09 \AA~ and 1.47$\pm$0.04 \AA~ respectively. The lines are not
resolved with the FOS and are likely to be saturated. GEA99
quote a CIV column density in the range between $6.3\times 10^{14}$ and
$10^{15}$ cm$^{-2}$, very similar to our predicted value for the warm
absorber. Earlier observations with GHRS have resolved the CIV
absorption line into four distinct components, two of which are from the
host galaxy of NGC~3516. The two nuclear components have column
densities consistent with the GEA99 values. We conclude that the
nuclear CIV absorption line (possibly component 1 of Crenshaw et al. (1998))
is likely to arise in the warm absorber observed by
BeppoSAX. At the least, the warm absorber makes a substantial
contribution to the absorption seen in the UV (based on CIV
strength). 

The hot X-ray absorber is too ionized to contribute to the
UV absorption lines, since carbon, oxygen and magnesium are 
fully ionized. 

\bigskip
X-ray and UV observations show that the system of X-ray absorbers 
in NGC~3516 is clearly complex and highly structured with multiple 
ionization stages just as in the UV absorption lines 
($\S 1$). 
This complexity is enhanced by variability on many different time 
scales, from days to years. 
It is possible that on the very short time scales of a day, the variability
of absorption is governed by the photoionization-recombination
timescales, while on the longer time scales from months to years the 
dynamical time may play the dominant role, either via bulk-motion 
or adiabatic expansion. 
It is conceivable that the different ionization stages trace different
phases in the evolution of an absorber. 
A neutral absorber may expand as it outflows, becomes a warm absorber 
and eventually becomes hot and later completely transparent in the X-rays, 
or the ``components'' may be part of a quasi-static structure in 
an outflowing wind (Elvis 2000, Murray \& Chiang 1995). 
Again high resolution Chandra and XMM observations of this source will 
be crucial to disentangle the many X-ray components, and clearly 
establish (or reject) a link with the UV components. 

\section{Conclusions}
We presented two BeppoSAX observations of NGC~3516, performed 
in 1996 November, and 1997 March. 

Our main findings are: 

\begin{itemize}

\item{} The source underwent strong spectral variability between 
the two BeppoSAX observations, taken 4 months apart, due entirely 
to the drastically different degree of obscuration of the 
central source. During the first observation the nuclear X-ray 
continuum was absorbed by an equivalent column of cold hydrogen of 
$2.3 \times 10^{22}\ cm^{-2}$, while in 1997 November the absorption was 
fully compatible with that from our Galaxy. 
This is the first time that such a change in the degree 
of obscuration of a Seyfert 1 has been clearly observed. 

We investigated several possibilities and propose 
as possible explanations either a BELC crossing the line 
of sight, or a state-varying absorber. 

\item{} We discovered the presence of a very highly ionized 
absorber/emitter in the nuclear environment of NGC~3516. 
This gas is visible in absorption (and with a stable physical 
state) in both BeppoSAX observations. 

Thanks to the large fraction of primary X-rays absorbed 
by neutral gas in 1996 we also were able to detect 
emission from this gas, and to give an estimate of its covering factor $f_{hot} = 0.3-0.4$

\end{itemize}

\acknowledgments 
We acknowledge partial support of the Italian Space Agency under contracts ASI-ARS-98-119
and ASI-ARS-99-15, of the Italian
Ministry for University and Research (MURST) under grant Cofin-98-02-32 and CXC
contract NAS8-39073. E.C. is supported by the NASA grant NAG5-3545. S.M.
aknowledges NASA grant NAG5-3249 LTSA. F.N. aknowledges NASA grant NAG5-9216.

\clearpage
\clearpage
\begin{deluxetable}{c ccc ccc}
\tablewidth{0pt}
\tablecaption{ Observation log of the two BeppoSAX pointings of
NGC~3516. Count rates are given in the bands 
0.1--2 keV, 2--10 keV and 15--150 keV, for the LECS, MECS and PDS 
respectively.}
\tablehead
{
\colhead{Dates} & \colhead {} & \colhead{Exposures} & \colhead{} &
\colhead {} & \colhead{Rate (cts/sec)} & \colhead {}\\
\colhead{} & \colhead{LECS} & \colhead{MECS} & \colhead{PDS} & \colhead{LECS} & \colhead{MECS} & \colhead{PDS}
}
\startdata
8-11-1996 & 16059 & 56013 & 20758 & 0.034$\pm$0.001 & 0.36$\pm$0.003 & 0.72$\pm$0.04\\
12-3-1997 & 16909 & 57712 & 21713 & 0.14$\pm$0.004 & 0.69$\pm$0.004 & 0.86$\pm$0.04\\
\enddata
\end{deluxetable}

\clearpage
\small
\begin{deluxetable}{cccccccccccccc}
\tabletypesize{\scriptsize}
\rotate
\tablewidth{0pt} 
\tablecaption{Best fitting parameterizations 96BF and 97BF. 
Both include: a cut-off power law plus Galactic absorption 
a gaussian line plus cold reflection and a highly ionized 
absorber. 96BF includes also a partial covering absorber (C$_{sou}\sim$0.95), and 
a thermal component. 97BF includes instead a ``warm absorber''.
See \S3 for details.}
\tablehead
{
\colhead{} & \colhead{$^{a}$Norm} & \colhead{$\Gamma$} & \colhead{$^b$N$_{H}$} 
& \colhead{$^c$kT} &
\colhead{$^d$log N$_{H}^{WA}$} & \colhead{log U$^{WA}$} & 
\colhead{$^c$E$_{K\alpha}$} & \colhead{$^c\sigma_{K\alpha}$} &
\colhead{$^e$EW$_{K\alpha}$} & \colhead{$^d$log N$_{H}^{HA}$} & 
\colhead{logU$^{HA}$}  & \colhead{R$^{f}$} & \colhead{$\chi^{2}$/dof}\\
}
\startdata
96BF & 1.43$\pm$0.13 & 2.03$^{+0.15}_{-0.17}$ & 2.31$^{+0.33}_{-0.15}$ &
1.78$^{+0.91}_{-0.39}$ & $\cdots$ & $\cdots$ &
6.33$\pm$0.30 & $<$0.52 & 108$^{+51}_{-45}$ & 23.5$\pm$0.1 &
2.35$\pm$0.2 & 1.2$^{+0.6}_{-0.2}$ & 91/94\\
\tableline
97BF & 2.38$\pm$0.05 & 2.04$^{+0.15}_{-0.16}$ &
2.77$^{+0.49}_{-0.44}\times$10$^{-2}$ & $\cdots$ & 22.09$\pm$0.02 & 0.73$\pm$0.
1 &
6.41$\pm$0.15 & $<$0.27 & 102$^{+71}_{-56}$ & 23.2$\pm$0.2 &
2.36$\pm$0.1 & 1.4$^{+0.2}_{-0.2}$ & 85/94\\
\enddata

$^a$in 10$^{-2}$
ph cm$^{-2}$s$^{-1}$keV$^{-1}$@1 keV. $^b$ In $10^{22}$ cm$^{-2}$. $^c$ In keV. $^d$ In cm$^{-2}$. $^e$ In eV .$^{f}$ errors are calculated
fixing the slope at the best fit value. 
\end{deluxetable}

\clearpage
\figcaption{Ratio of the N96 and M97 spectra of  NGC~3516.}

\figcaption{Residuals of a simple power law plus Galactic
absorption fit for N96 (upper panel) and M97 (lower panel).} 

\figcaption{Upper panel: N96 residuals of the ``base'' model 
plus a partial covering neutral absorber; an emission feature 
at 1 keV is clearly present. Lower panel: M97 residuals of the ``base'' 
model; the typical warm absorber signature is visible.}

\figcaption{96BF model with no CIP component (upper panel) and 
97BF model (lower panel). Photoionization models include gas emission 
and resonant absorption other than photoelectric absorption 
(see text for details).}
\figcaption{Ratios between the two models in the upper 
and lower panels, and the raw 0.1-10 keV N96 and M97 spectra.}

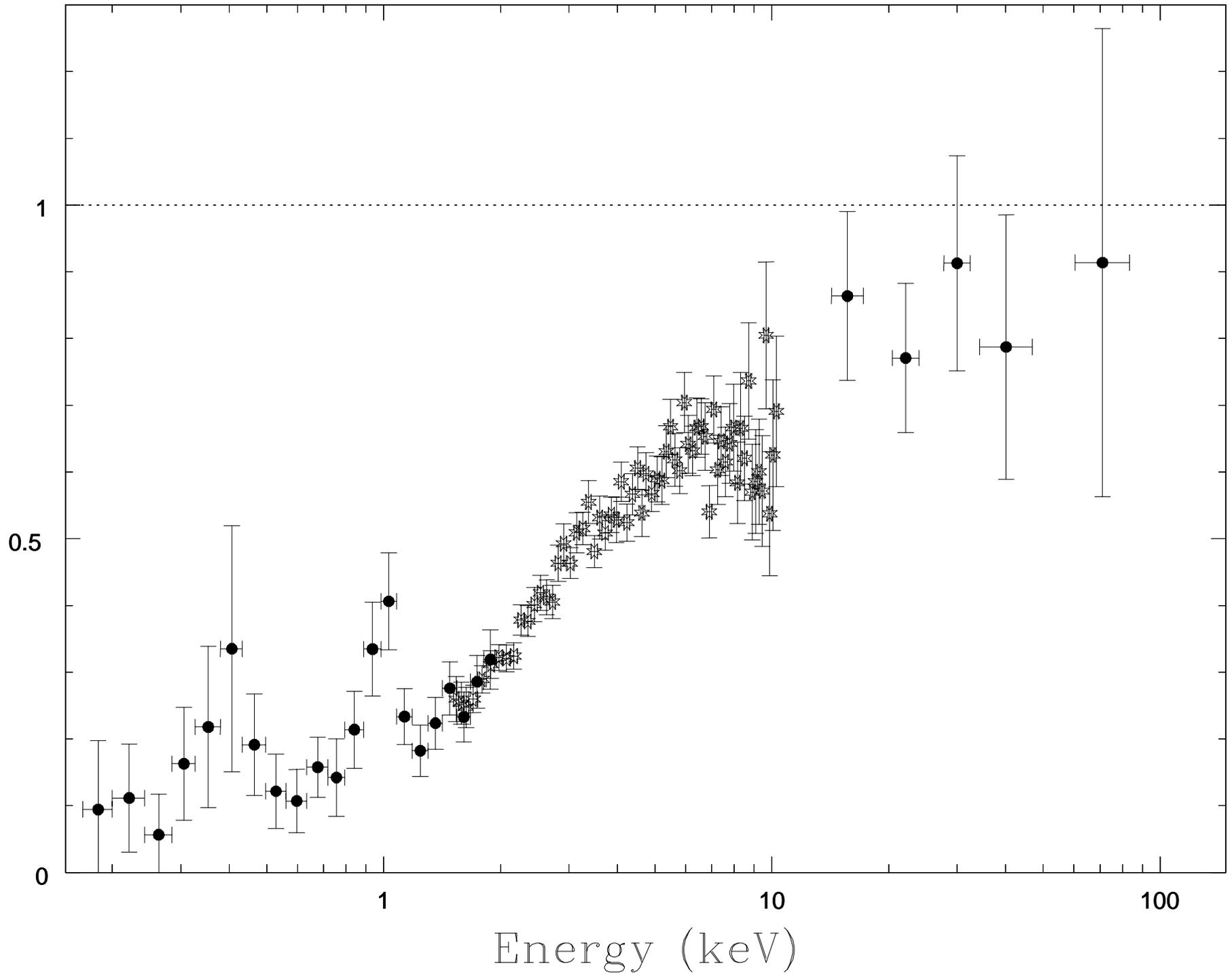
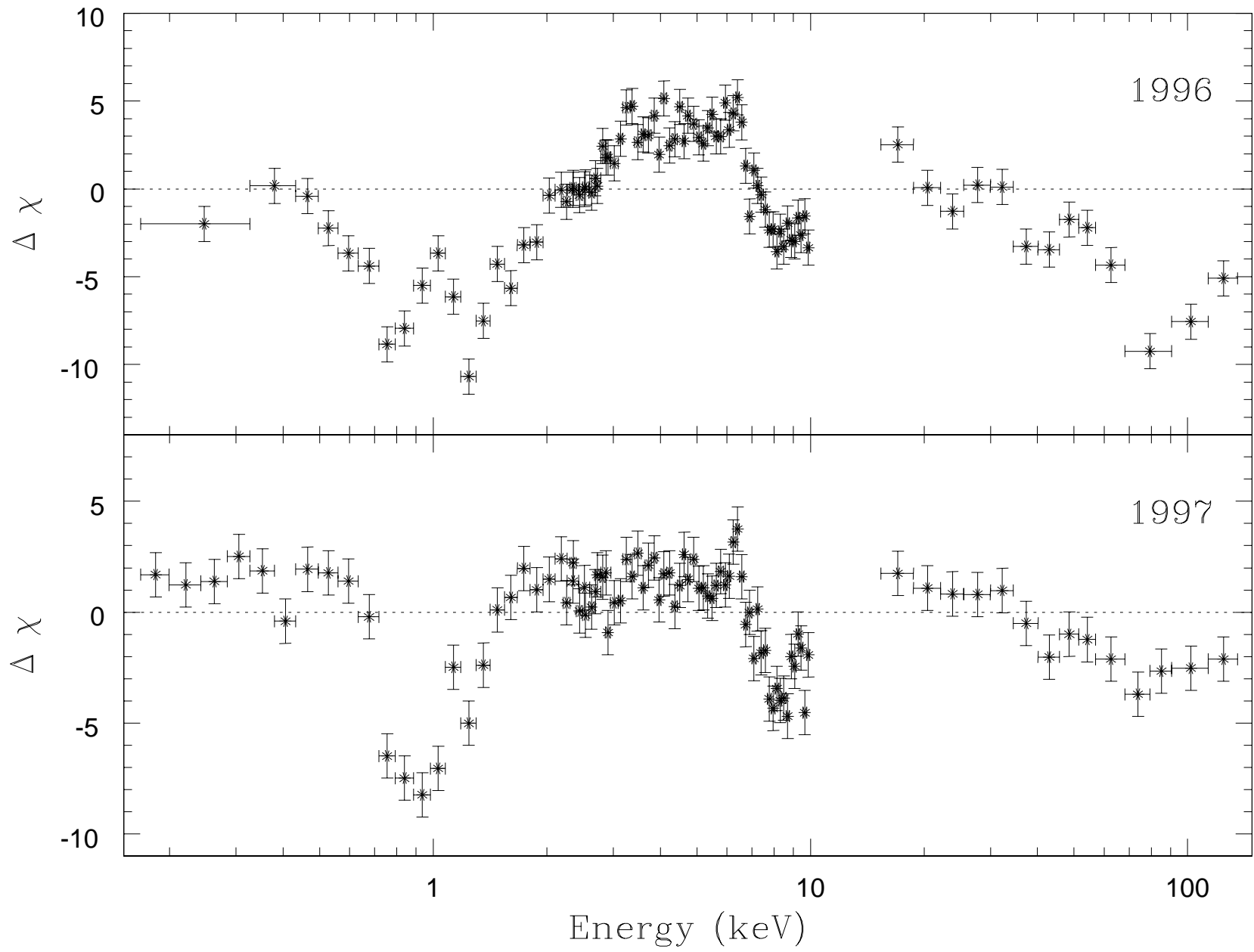
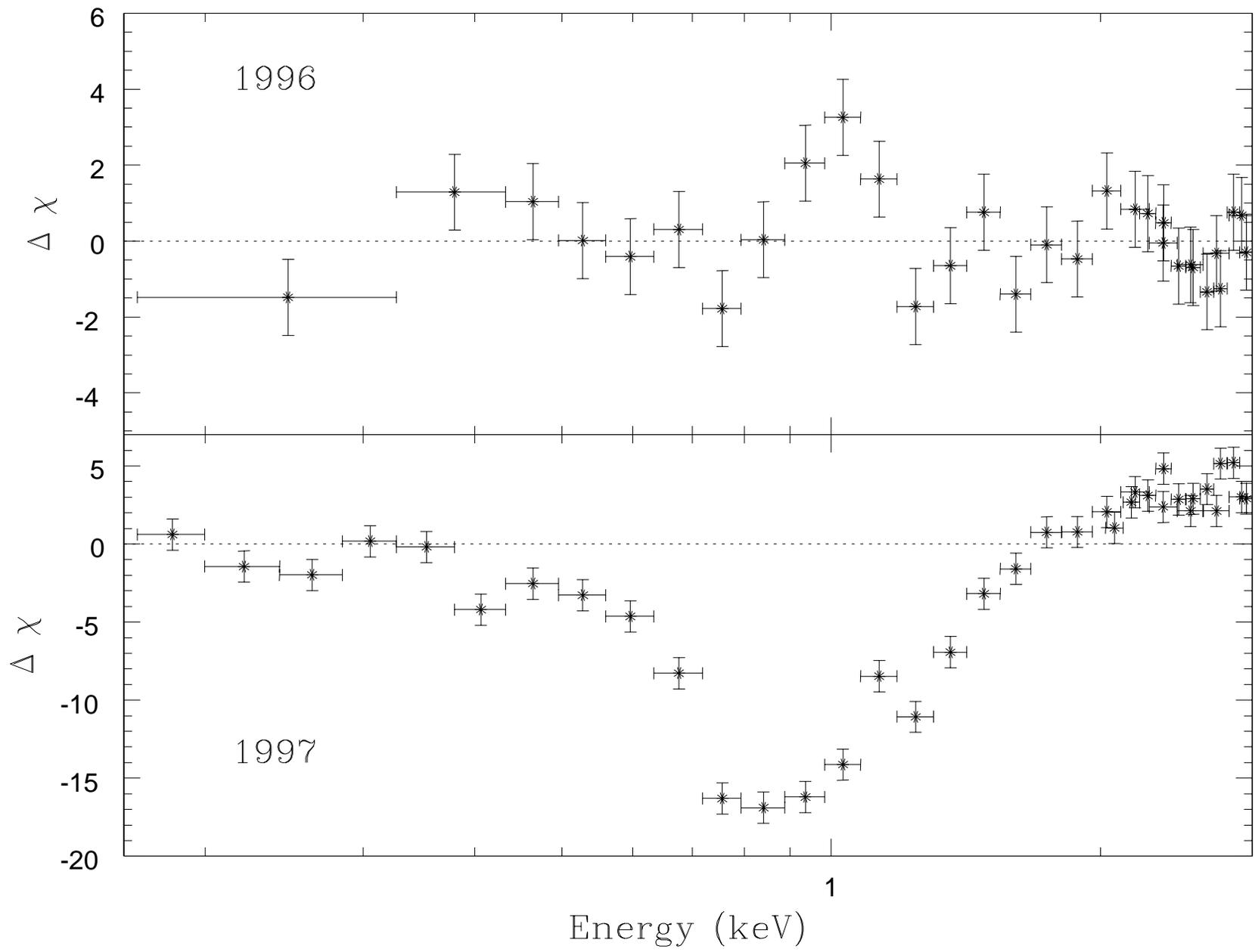
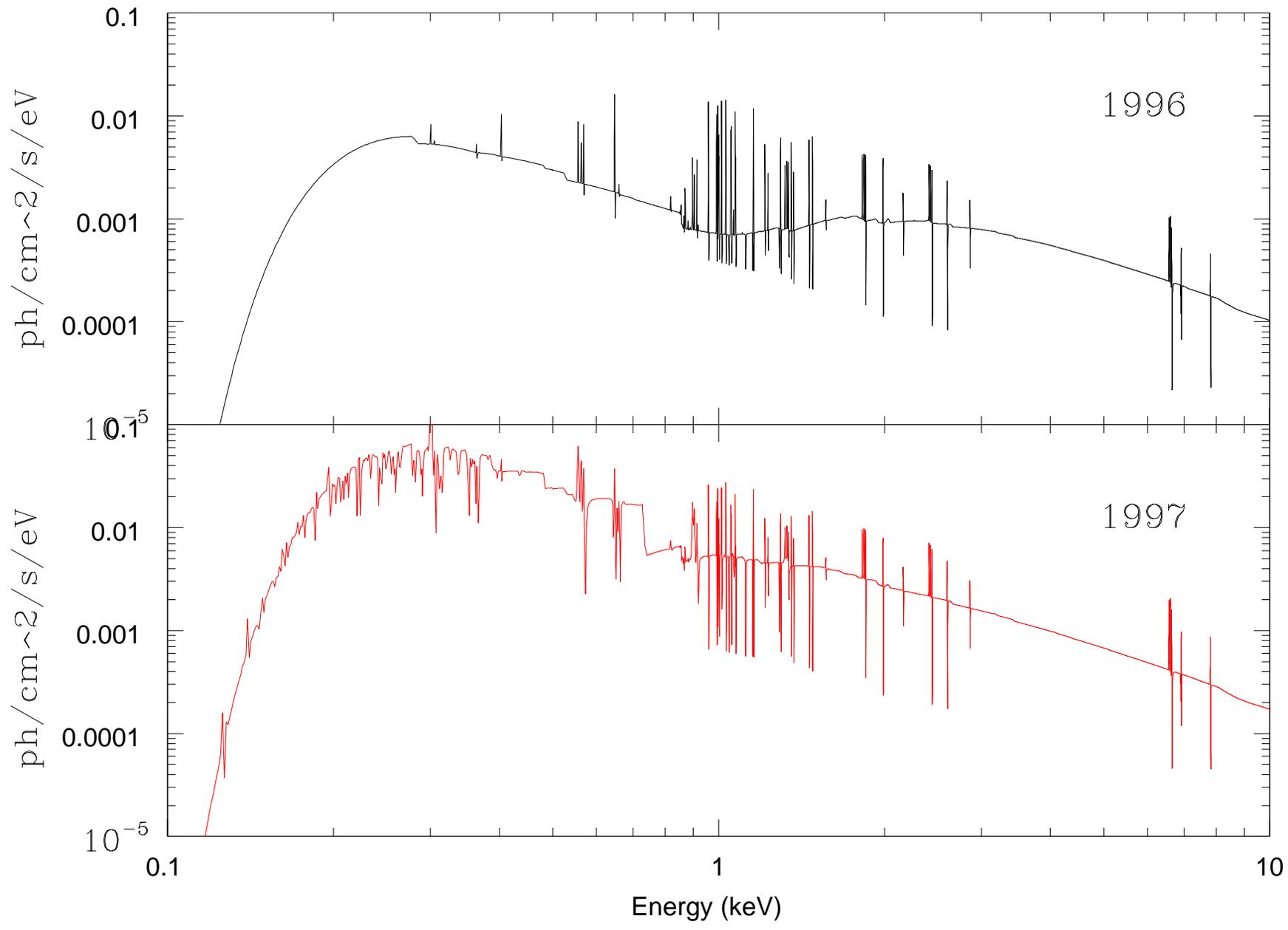
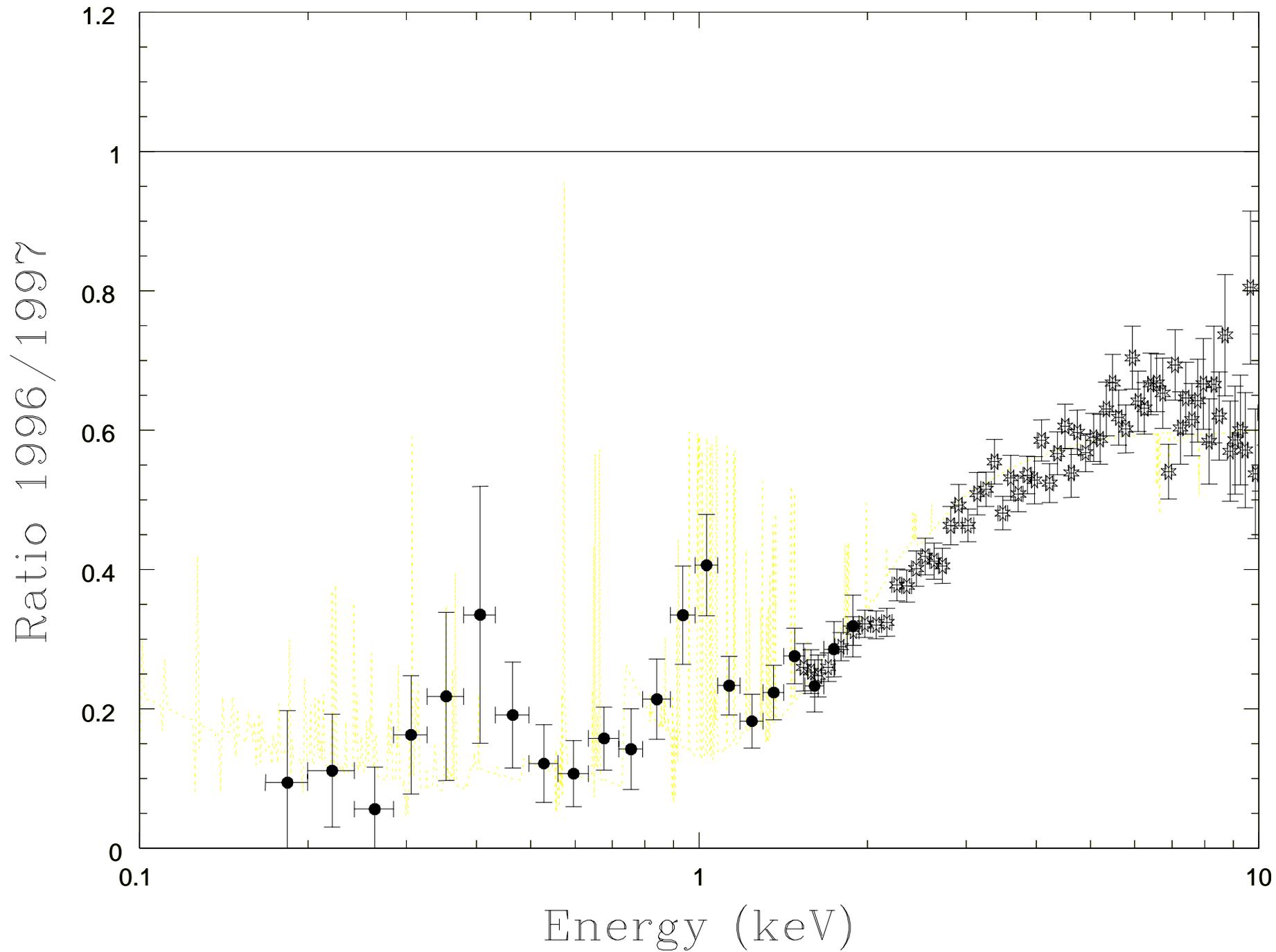
\end{document}